# Structure Re-Determination and Superconductivity Observation of Bulk 1T $MoS_2$

Yuqiang Fang[abc§], Jie Pan[a§], Jianqiao He[abc], Ruichun Luo[d], Dong Wang[a], Xiangli Che[a], Kejun Bu[abc], Wei Zhao[a], Pan Liu[d], Gang Mu[e*], Hui Zhang[e], Tianquan Lin[a]*, Fuqiang Huang[ac]*

**2H $MoS_2$ has been intensively studied because of layer-dependent electronic structures and novel physical properties. Though the metastable 1T $MoS_2$ with the [$MoS_6$] octahedron was observed from the microscopic area, the true crystal structure of 1T phase has not been determined strictly. Moreover, the true physical properties have not been demonstrated from experiments due to the challenge for the preparation of pure 1T $MoS_2$ crystals. Here, we successfully synthesized the 1T $MoS_2$ single crystals and re-determined the crystal structure of 1T $MoS_2$ from single-crystal X-ray diffraction. 1T $MoS_2$ crystalizes in space group P-3m1 with a cell of *a* = *b* = 3.190(3) Å and *c* = 5.945(6) Å. The individual $MoS_2$ layer consists of $MoS_6$ octahedron sharing edge with each other. More surprisingly, the bulk 1T $MoS_2$ crystals undergo a superconducting transition of $T_c$ = 4 K, which is the first observation of superconductivity in pure 1T $MoS_2$ phase.**

Two-dimensional crystals $MoS_2$ have attracted growing interests in recent years due to abundant phase evolutions and novel electronic structures.[1] Depending on different coordination configurations of [$MoS_6$] polyhedral, $MoS_2$ can crystalize in different phases including 2H, 1T, 1T', 1T'', *etc*. 2H $MoS_2$ is a thermodynamically stable phase containing A-b-A sandwich-like layers with edge-sharing [$MoS_6$] trigonal prisms (see **Figure 1a**).[2] The metastable 1T $MoS_2$ was reported to have A-b-C layers with edge-sharing [$MoS_6$] octahedron (see **Figure 1b**), which was derived from electron diffraction[3] and powder X-ray pattern.[4] But the crystal structure of 1T $MoS_2$ has never been collected in the Inorganic Crystal Structure Database (ICSD) due to the lack of a strict structural refinement.

Although the single-layer 1T $MoS_2$ has been prepared by using various synthetic methods,[5] the obtained nanosheets always coexist with 2H, 1T, and 1T' phases. The maximal content of 1T/1T' domains have been reported to be up to 80% which was evaluated by XPS measurements. [5b] The 1T/1T' $MoS_2$ phases were observed to start the transformation

[a]State Key Laboratory of High Performance Ceramics and Superfine Microstructure, Shanghai Institute of Ceramics, Chinese Academy of Sciences, Dingxi Road, 1295, Shanghai (P. R. China). [b]University of Chinese Academy of Sciences, Yuquan Road, 19, Beijing (P. R. China). [c]State Key Laboratory of Rare Earth Materials Chemistry and Applications, College of Chemistry and Molecular Engineering, Peking University, Chengfu Road, 202, Beijing (P. R. China). [d]State Key Laboratory of Metal Matrix Composites, School of Materials Science and Engineering, Dongchuani Road, 800, Shanghai (P. R. China), Shanghai Jiao Tong University. [e]State Key Laboratory of Functional Materials for Informatics, Shanghai Institute of Microsystem and Information Technology, Chinese Academy of Sciences, Changning Road, 865, Shanghai (P. R. China)

E-mail: lintianquan@mail.sic.ac.cn; mugang@mail.sim.ac.cn; huangfq@mail.sic.ac.cn

[§]These authors contributed equally to this work.

to thermodynamically more stable 2H phase at around 100 °C and be completely converted into 2H phase at 300 °C.[6] It was recently reported that 1T/1T' $MoS_2$ domains can be stabilized by electron doping.[7]

2H $MoS_2$ intrinsically behaves as a semiconductor with a band gap of 1.2-1.9 eV,[8] which is consistent with the close-shell electronic configuration of $Mo^{4+}$ $4d^2$ ($d_{z^2}$, $d_{x^2-y^2}$) in a trigonal prismatic coordination environment (**Figure 1a**).[9] Previously reported superconductivity in the $MoS_2$ system is based on electron injection to the Fermi surface of 2H $MoS_2$ through physical and chemical routes. When alkali metal atoms are intercalated into the [$MoS_2$] interlayers, partial electronic states of Mo $4d_{x^2-y^2}$ and $4d_{xy}$ are filled to generate itinerant electrons, which cause the superconductivity with transition temperature ($T_c$) ~3-6.5 K.[10] More recently, a maximum $T_c$ up to 10.8 K was observed in the ionic liquid gated 2H-$MoS_2$.[11] $Mo^{4+}$ $4d^2$ orbitals in an octahedral crystal field of 1T $MoS_2$ are split into the well-known configuration of $e$ ($d_{x^2-y^2}$, $d_{z^2}$) over $t$ ($d_{xy}$, $d_{xz}$, $d_{yz}$)[9] shown in **Figure 1b**, whose two electrons are filled in the $t$ states and may also become itinerant electrons. Anomalous superconductivity has recently been reported in the structure-related 1T' $MoTe_2$ at a very low $T_c$ of about 0.1 K.[12] The density functional calculations predicted the metallicity of 1T $MoS_2$.[13] However, the sample of $MoS_2$ nanosheets coexisting with 1T/1T' and 2H phases was found to have a semiconducting behavior.[14] Moreover, the superconductivity of 1T $MoS_2$ has never been discovered yet.

So far, it is very important and urgent to synthesize single crystals of 1T-$MoS_2$ for re-determining single crystal structure and investigating intrinsic physical properties. Here, we reported a modified strategy derived from the literature[4] to prepare metastable 1T-$MoS_2$ single crystals. The crystal structure of 1T-$MoS_2$ was determined by single crystal X-ray diffraction for the first time. The 1T-$MoS_2$ crystal structures are proved through scan transmission electron microscopy (STEM), and Raman spectrum. The superconductivity of 1T $MoS_2$ at 4 K was observed for the first time. The zero temperature upper critical magnetic field of 1T-$MoS_2$ is estimated to be 5.02 T based on the temperature dependence of the upper critical magnetic field.

The preparation of 1T-$MoS_2$ started from the intercalated compound $LiMoS_2$, whose synthesis process was described by our previous results.[15] The oxidation processes of $LiMoS_2$ crystals were divided into two steps, in which the corresponding reactions happen as follow:

$$LiMoS_2 + xH_2O \rightarrow Li_{1-x}(H_2O)_yMoS_2 + xLiOH + x/2H_2(g) \quad (1)$$

$$Li_{1-x}(H_2O)_yMoS_2 + (1-x)/2\ I_2 \rightarrow MoS_2 + (1-x)\ LiI + yH_2O \quad (2)$$

In the first step, lithium atoms undergo the hydration process, which expands the interlayer distance of $MoS_2$. **Figure S1a** shows the PXRD pattern of as-obtained $Li_{1-x}(H_2O)_yMoS_2$, which corresponds to 12.149 Å of the $d$ value along the $c$ direction. Expanded layer spacing reduces the interactions between sulfur atoms and lithium atoms, which promotes the removal of lithium ion in the second redox process. Moreover, the deintercalation of Li ions causes fewer damages on the crystal structure of 1T $MoS_2$ compared with that caused by the deintercalation of K ions.[3],[4]

As illustrated in **Figure 1b**, the 1T-$MoS_2$ crystallizes in the trigonal space group $P$-3$m$1. The unit cell consists of one independent Mo site and one independent S site. Each layer is composed of $[MoS_6]^{8-}$ octahedron interconnected via edge sharing along $ab$ plane. The crystal data for 1T-$MoS_2$ are summarized in **Table S1.** And the refinement details are listed in **Supplemental Information (Table S2-S4).** During the delicate deintercalation of Li ions, the occurrence of some tiny distortions and local stress could weaken the lattice completeness of the 1T $MoS_2$ single crystal, leading to increased $R$-factor values. Moreover, the relatively large $U_{33}$ value of the S atom reflect the thermal motion along the $c$ axis in 1T $MoS_2$, which could arises from the interactions between interlayers.

The chemical components of the sample were analyzed by the EDX and ICP-AES measurements. The EDX data (**Table S5**) indicate the atomic ratio of Mo and S is close to 1:2. The result of ICP-AES measurement demonstrates that the atomic fraction of $Li^+$ ions in the sample is around 0.18‰ (the atomic ratio is Li:Mo:S=0.00054:1:1.986), which could arise from the adsorption of LiI or LiOH on sample surface. Therefore, the influence of $Li^+$ doping on the superconductivity of the bulk 1T $MoS_2$ can be neglected. The powder X-ray diffraction (PXRD) pattern of the obtained $MoS_2$ sample is shown in **Figure 2a**. The measured pattern of the compound matches well with the simulated one obtained from single-crystal data except for some low-intensity reflections in the range of 47° to 55°. However, none of reactants or by-products can match with these extra peaks, which belong to some unknown compounds. The Rietveld-Refinement of the PXRD data was conducted, shown in **Figure S2**. The relatively low $R_p$ and $wR_p$ demonstrate that the PXRD data fit well with the single-crystal diffraction data of 1T $MoS_2$ except for some undefined diffraction peaks. Besides, the obvious preferred orientation demonstrates the plate-like texture, consistent with the result of the SEM measurements shown in **Figure S1b**. The Raman spectrum of the sample measured at room temperature is shown in **Figure 2b**. There are five peaks located at 150 $cm^{-1}$, 212 $cm^{-1}$, 280 $cm^{-1}$, 324 $cm^{-1}$ and 405 $cm^{-1}$, respectively, which corresponds to the $J_1$, $J_2$, $E_{1g}$, $J_3$ and $A_{1g}$ modes in 1T $MoS_2$.[16] Specially, the $E_{1g}$ peak at 280 $cm^{-1}$ arises from the octahedral coordination of Mo atoms in 1T $MoS_2$. The disappearance of the characteristic Raman mode in 2H-$MoS_2$ situated at 383 $cm^{-1}$($E_{2g}^1$ mode) demonstrate the absence of 2H-$MoS_2$ from the macroscopic scale.

The microscopic structural evidence of the 1T-$MoS_2$ crystals was observed by the high-angle annular dark-field scanning transmission electron microscopy (HAADF-STEM). As shown in **Figure 3a**, a hexagonal lattice can be observed. Moreover, one Mo atom is surrounded by six nearest-neighbor Mo atoms, which is the direct evidence of the 1T phase.[17] **Figure 3b** presents the difference of Mo atom arrangement between the bulk 1T phase and the 2H phase. In bulk 2H $MoS_2$, each Mo atom is surrounded by three nearest neighboring Mo atoms. Besides, the Mo-Mo distance in the **Figure 3a** is 3.39 Å, consistent with the value obtained by the single X-ray diffraction data. Selected area electron diffraction (SEAD) of 1T-$MoS_2$ crystal in the inset of **Figure 3a** exhibits the diffraction spots of (100) and (-120) planes, whose *d* values are 2.80 Å and 1.62 Å respectively.

Subsequently, we investigated the intrinsic physical properties of 1T-$MoS_2$ crystals. **Figure 4a** shows the temperature-dependent magnetic susceptibility under a magnetic field of 40 Oe. The onset diamagnetic transition for the 1T-$MoS_2$ crystals is 4 K. And the superconducting volume fraction is around 22% at 2 K. Furthermore, the specific heat capacity of bulk 1T $MoS_2$ displays an abnormal change around 4.2 K (**Figure S3**), which arises from the superconducting transition. The transition temperature of 4.2 K is consistent with the results of the magnetic measurement. Both of the magnetic and specific heat capacity measurements confirm the superconductivity of bulk 1T $MoS_2$. Furthermore, the profile of the magnetization-field curve of 1T-$MoS_2$ at 2 K (inset of **Figure 4a**) proves that 1T-$MoS_2$ is a typical type-II superconductor.[18] **Figure 4b** shows the temperature-dependent electrical resistivity of 1T-$MoS_2$ crystals, measured on a cold-pressed pellet through the four-probe method. Before the superconducting transition, the resistivity of the sample rises as the temperature decreases, which performs the semiconducting-like behavior. Subsequently the resistivity starts undergoing an abrupt decrease at around 5 K, and reaches zero at 3 K.

The electrical propertie of $Li_x(H_2O)_yMoS_2$ crystals were also measured, which shows the typical semiconducting behavior without the superconducting transition (**Figure S3**). In order to define the upper critical field of 1T-$MoS_2$ crystals, the temperature-dependent resistivity of 1T-$MoS_2$ was measured under a series of magnetic fields. Superconducting transition temperature $T_c$ was taken as the temperature where the resistivity drops to 90% of the normal resistivity (**Figure 4c**). The $T_c$ decreases as the magnetic field increases. **Figure 4d** shows the temperature dependence of the upper critical magnetic field. The upper critical magnetic field at T = 0 K can be obtained by the Werthamer –Helfand –Hohenberg (WHH) formula $H_{c2}(0) = 0.693[-(dH_{c2}/dT)]_{Tc}T_c$ according to the BCS theory.[19] The value of $dH_{c2}/dT$ is -1.55 T $K^{-1}$ and the estimated $H_{c2}(0)$ is 5.02 T.

In summary, we successfully prepared the pure 1T $MoS_2$ single crystals, and the crystal structure of 1T-$MoS_2$ was determined by the single-crystal diffraction for the first time. From the magnetic and electrical measurements, the 1T-$MoS_2$ crystals were observed to be a superconductor under 4 K, whose upper critical field $H_{c2}$ is 5.02 T. Our finding is the first report on the intrinsic superconductivity in the 1T $MoS_2$ phase, compared with the reported superconductivity of the 2H $MoS_2$ induced by external electron doping through intercalation of alkali metal ions or liquid ionic gating technology.

# Accession Codes

CCDC 1583585 contain the supplementary crystallographic data for this paper. These data can be obtained free of charge via www.ccdc.cam.ac.uk/data_request/cif, or by emailing data_request@ccdc.cam.ac.uk, or by containing The Cambridge Crystallographic Data Centre, 12 Union Road, Cambridge CB2 1EZ, UK; fax: +44 1223 336033.

# Acknowledgements

This work was financially supported by National key R&D Program of China (Grant 2016YFB0901600), Science and Technology Commission of Shanghai (Grant 16JC1401700), National Science Foundation of China (Grant 51672301), the Key Research Program of Chinese Academy of Sciences (Grants No. QYZDJ-SSW-JSC013 and KGZD-EW-T06), Innovation Project of Shanghai Institute of Ceramics (Grant No. Y73ZC6160G), and CAS Center for Excellence in Superconducting Electronics.

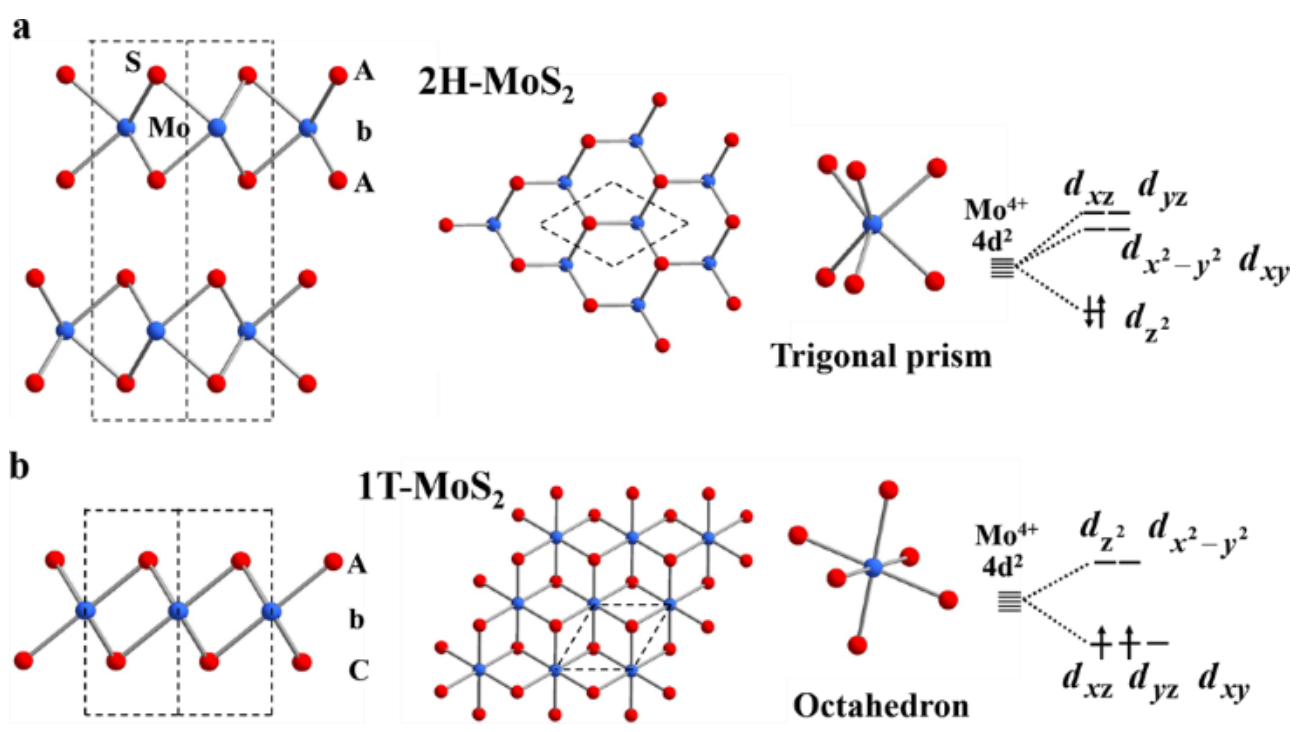

**Figure 1. a,** the schematic diagram of the 2H-$MoS_2$ crystal structure and the ligand splitting for the 4d orbitals of Mo atoms with trigonal-prismatic coordination. **b,** the schematic diagram of the crystal structure of 1T $MoS_2$ and the ligand splitting for 4d orbitals of Mo atoms with octahedral coordination.

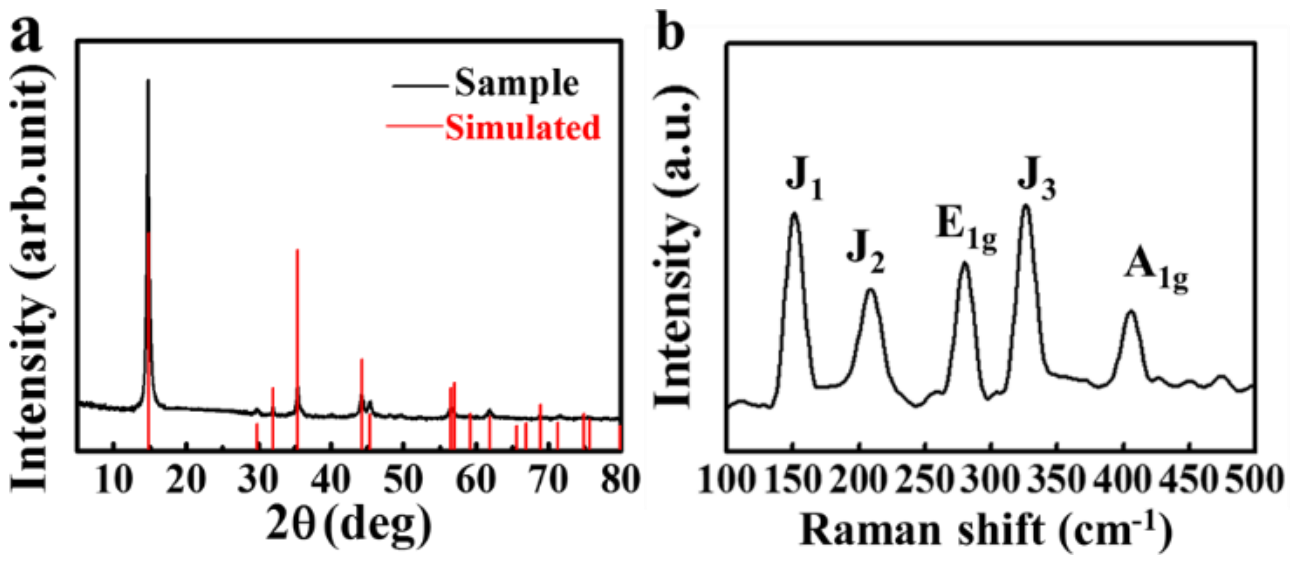

**Figure 2. a,** Powder X-ray diffraction patterns of the obtained 1T $MoS_2$ powder sample, where the red lines were calculated from single crystal structure of 1T $MoS_2$. **b,** Raman spectrum of 1T $MoS_2$ powder sample.

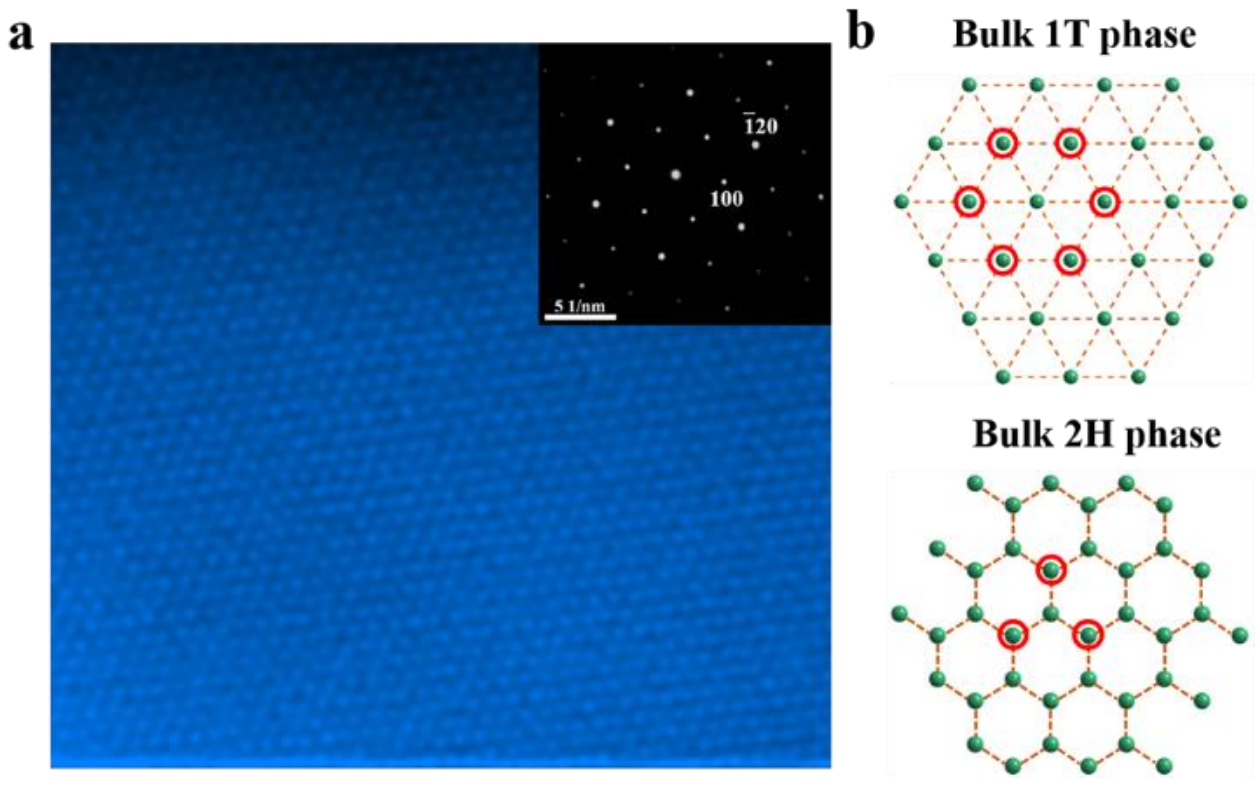

**Figure 3. a,** The HAADF-STEM image of 1T-$MoS_2$, inset: selected area electron diffraction (SAED) of 1T-$MoS_2$ crystal. **b,** Schematic arrangements of Mo atoms presenting six neighboring Mo atoms in bulk 1T-$MoS_2$ and three neighboring Mo in bulk 2H-$MoS_2$, respectively.

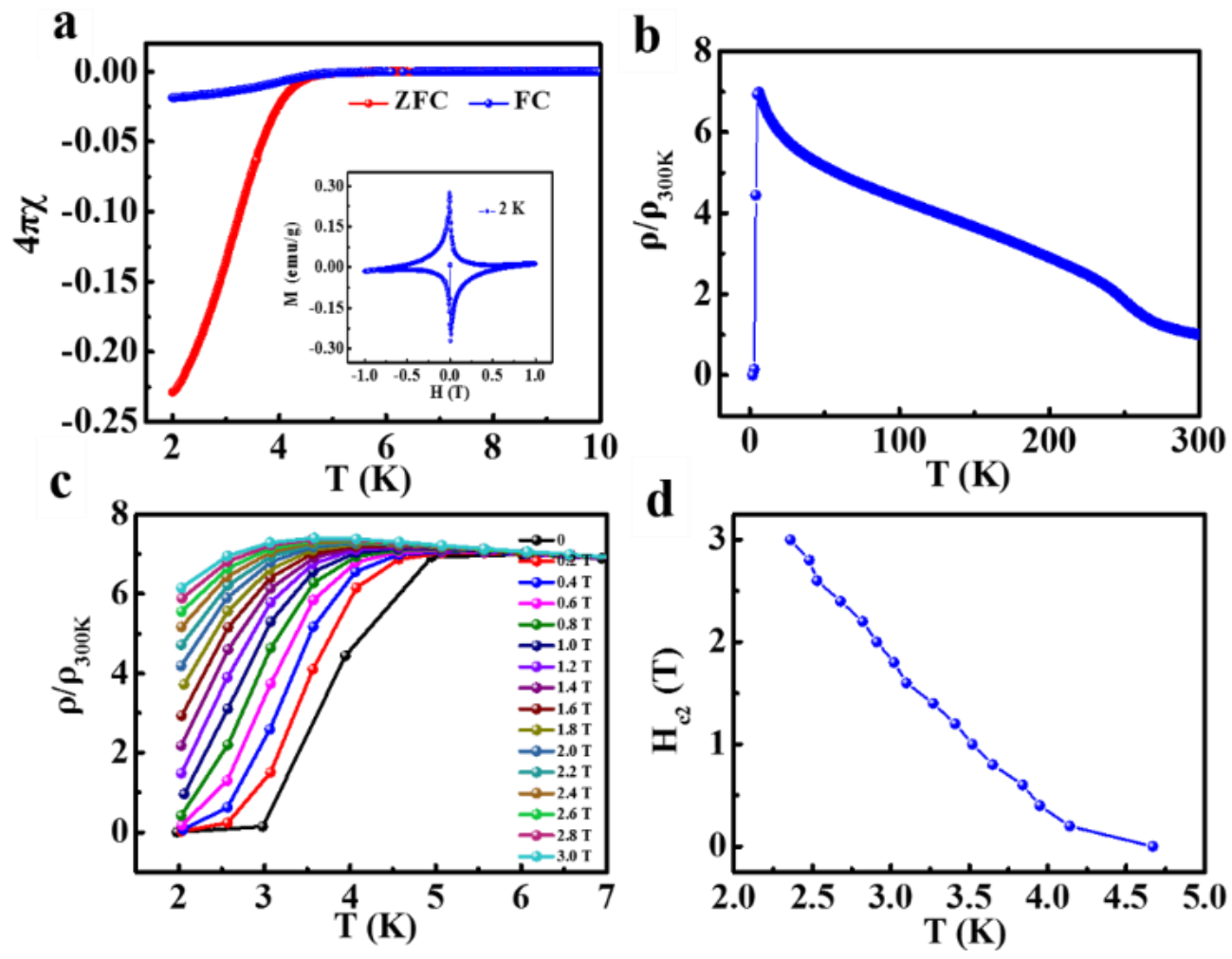

**Figure 4. a,** Temperature dependence of magnetic susceptibility of 1T $MoS_2$ under the magnetic field of 40 Oe, with zero-field-cooling curve (red) and field-cooling curve (blue). The inset shows the magnetic hysteresis of the sample measured at 2K. **b,** Temperature dependence of electrical resistivity of 1T-$MoS_2$ **c,** Electrical resistivity of 1T-$MoS_2$ measured under magnetic fields of 0, 0.2, 0.4, 0.6, 0.8, 1, 1.2, 1.4, 1.6, 1.8, 2, 2.2, 2.4, 2.6, 2.8, 3 T below 7 K. **d,** The relationship between $T_c$ and magnetic field.

*Supporting Information*

**List of contents:**

## 1. Materials and Methods

### Preparation of $LiMoS_2$

The $LiMoS_2$ crystals were synthesized *via* the high temperature solid-state reactions. Experimental operations were taken inside an argon-filled glove box, $Li_2S$ (99.9%, Alfa Aesar), Mo (99.9%, Alfa Aesar), and S (99.9%, Alfa Aesar) were mixed in the ratio of 1:2:3, then they were ground in a mortar. The mixed powders were pressed into a pellet and transferred into a quartz tube. The tube was evacuated $<10^{-5}$ Torr and sealed. Subsequently the quartz tube was held at 850 $^0$C for 20 h, followed by naturally cooling to room temperature.

### Preparation of 1T $MoS_2$

The synthesised $LiMoS_2$ crystals were soaked in water and immediately they reacted with water to generate $H_2$ gas. When there is no emergence of $H_2$ gas, the reaction ended and the $Li_{1-x}(H_2O)_yMoS_2$ samples were obtained. Then the $Li_{1-x}(H_2O)_yMoS_2$ samples were filtrated and dried in a vacuum oven. Subsequetly the $Li_{1-x}(H_2O)_yMoS_2$ samples were stirred in the 50mL of a 0.04mol/L acetonitrile solution of $I_2$ for 30 minutes. Finally, residual $I_2$ and the byproducts LiI were removed by alcohol. After dried in the vacuum oven, the 1T $MoS_2$ crystals were obtained.

### Single crystal determination

Suitable crystals were chosen to perform the data collections. Single-crystal X-ray diffraction was performed on a Bruker D8QUEST diffractometer equipped with Mo K $\alpha$ radiation. The diffraction data were collected at room temperature by the $\omega$- and $\varphi$-scan methods. The crystal structures were solved and re fi ned using APEX3 program. Absorption corrections were performed using the multiscan method (ASDABS).The detailed crystal data and structure re finement parameters are summarized in **Table 1**.

**Superconducting measurement**

Temperature-dependent magnetic suscepitibility measurement was conducted under a magnetic field of 40 Oe, containg field-cooling and zero field-cooling processes. The magnetic field-dependent magnetic susceptibility measurement was taken at 2 K, and the resisity measurement as the function of temperature was conducted under different magnetic fields. Abovementioned measurements were performed in a physical property measurement system (PPMS) of Quantum design.

## 2. Supplementary Tables and Figures

**Table S1. Crystallographic Data for 1T-$MoS_2$.**

| atom | x | y | z | Wyckoff. | occ. |
|---|---|---|---|---|---|
| Mo | 0 | 0 | 0 | 1*a* | 1 |
| S | 1/3 | 2/3 | 0.2488(7) | 2*d* | 1 |

Space group *P*-3*m*1 (No. 147), *a* = *b* = 3.190(3) Å, *c* = 5.945(6) Å, Z = 1, and V =52.39(12) Å$^3$.

**Table S2. Crystal Data and Structural Refinement statistics for 1T $MoS_2$.**

| **formula** | **$MoS_2$** |
|---|---|
| $F_w$ (g mol$^{-1}$) | 160.06 |
| crystal system | trigonal |
| space group (No.) | *P*-3*m*1 (164) |
| *a* (Å) | 3.190 (3) |
| *b* (Å) | 3.190 (3) |
| *c* (Å) | 5.945 (6) |
| $\alpha$ (°) | 90 |
| $\beta$ (°) | 90 |
| $\gamma$ (°) | 120 |
| *V* (Å$^3$) | 52.39 (12) |
| *Z* | 1 |
| $\rho_c$ (g·cm$^{-3}$) | 5.073 |
| $\mu$ (mm$^{-1}$) | 7.50 |
| $\lambda$ (Mo K$\alpha$) (Å) | 0.71073 |
| *T* (K) | 293 |
| *F*(000) | 74 |
| $\theta_{max}$ (°) / completeness (%) | 24.975 / 96.2 |
| $R_{int}$ | 0.1418 |
| $R_1$[a]$[F^2 > 2\sigma(F^2)]$ | 0.057 |
| $wR_2$[b]$(F^2)$ | 0.1197 |
| goodness of fit | 1.222 |
| largest diff. peak and hole (e/Å$^3$) | 1.12 and -0.74 |

[a]$R_1 = \sum||F_o| - |F_c||/\sum|F_o|$. [b]$wR_2=\{\sum[w(F_o^2 - F_c^2)^2]/\sum[w(F_o^2)^2]\}^{1/2}$, $w = 1/[\sigma^2(F_o^2) + (0.1000P)^2]$ where $P = (F_o^2 + 2F_c^2)/3$.

**Table S3**. Anisotropic displacement parameters (Å$^2$) for 1T $MoS_2$.

| atom | $U_{11}$ | $U_{22}$ | $U_{33}$ | $U_{12}$ | $U_{13}$ | $U_{23}$ |
|---|---|---|---|---|---|---|
| Mo | 0.042 (2) | 0.042 (2) | 0.031 (2) | 0.0212 (11) | 0 | 0 |
| S | 0.020 (3) | 0.020 (3) | 0.065 (5) | 0.0099 (14) | 0 | 0 |

The anisotropic displacement factor exponent takes the form: $-2\pi^2[h^2a^{*2}U_{11} + ... + 2hka^*b^*U_{12}]$.

**Table S4**. Selected bond distances (Å) and bond angles (º) for 1T $MoS_2$.

| | bond distance (Å) | | bond distance (Å) |
|---|---|---|---|
| Mo—$S^{i}$ | 2.389 (5) | Mo—$Mo^{vi}$ | 3.190 (3) |
| Mo—$S^{ii}$ | 2.389 (5) | Mo—$Mo^{v}$ | 3.190 (3) |
| Mo—$S^{iii}$ | 2.389 (5) | Mo—$Mo^{vii}$ | 3.190 (3) |
| Mo—S | 2.389 (5) | Mo—$Mo^{viii}$ | 3.190 (3) |
| Mo—$S^{iv}$ | 2.389 (5) | Mo—$Mo^{ix}$ | 3.190 (3) |
| Mo—$S^{v}$ | 2.389 (5) | S—$Mo^{vii}$ | 2.389 (5) |
| Mo—$Mo^{iii}$ | 3.190 (3) | S—$Mo^{vi}$ | 2.389 (5) |
| | bond angle (°) | | bond angle (°) |
| $S^{i}$—Mo—$S^{ii}$ | 83.8 (2) | $Mo^{vi}$—Mo—$Mo^{v}$ | 120.0 |
| $S^{i}$—Mo—$S^{iii}$ | 96.2 (2) | $S^{i}$—Mo—$Mo^{vii}$ | 131.89 (11) |
| $S^{ii}$—Mo—$S^{iii}$ | 180.0 | $S^{ii}$—Mo—$Mo^{vii}$ | 90.0 |
| $S^{i}$—Mo—S | 180.0 | $S^{iii}$—Mo—$Mo^{vii}$ | 90.0 |
| $S^{ii}$—Mo—S | 96.2 (2) | S—Mo—$Mo^{vii}$ | 48.11 (11) |
| $S^{iii}$—Mo—S | 83.8 (2) | $S^{iv}$—Mo—$Mo^{vii}$ | 48.11 (11) |
| $S^{i}$—Mo—$S^{iv}$ | 83.8 (2) | $S^{v}$—Mo—$Mo^{vii}$ | 131.89 (11) |
| $S^{ii}$—Mo—$S^{iv}$ | 83.8 (2) | $Mo^{iii}$—Mo—$Mo^{vii}$ | 120.0 |
| $S^{iii}$—Mo—$S^{iv}$ | 96.2 (2) | $Mo^{vi}$—Mo—$Mo^{vii}$ | 60.0 |
| S—Mo—$S^{iv}$ | 96.2 (2) | $Mo^{v}$—Mo—$Mo^{vii}$ | 180.0 |
| $S^{i}$—Mo—$S^{v}$ | 96.2 (2) | $S^{i}$—Mo—$Mo^{viii}$ | 90.0 |
| $S^{ii}$—Mo—$S^{v}$ | 96.2 (2) | $S^{ii}$—Mo—$Mo^{viii}$ | 131.89 (11) |
| $S^{iii}$—Mo—$S^{v}$ | 83.8 (2) | $S^{iii}$—Mo—$Mo^{viii}$ | 48.11 (11) |
| S—Mo—$S^{v}$ | 83.8 (2) | S—Mo—$Mo^{viii}$ | 90.0 |
| $S^{iv}$—Mo—$S^{v}$ | 180.0 | $S^{iv}$—Mo—$Mo^{viii}$ | 48.11 (11) |
| $S^{i}$—Mo—$Mo^{iii}$ | 48.11 (11) | $S^{v}$—Mo—$Mo^{viii}$ | 131.89 (11) |
| $S^{ii}$—Mo—$Mo^{iii}$ | 131.89 (11) | $Mo^{iii}$—Mo—$Mo^{viii}$ | 60.0 |
| $S^{iii}$—Mo—$Mo^{iii}$ | 48.11 (11) | $Mo^{vi}$—Mo—$Mo^{viii}$ | 120.0 |
| S—Mo—$Mo^{iii}$ | 131.89 (11) | $Mo^{v}$—Mo—$Mo^{viii}$ | 120.0 |
| $S^{iv}$—Mo—$Mo^{iii}$ | 90.0 | $Mo^{vii}$—Mo—$Mo^{viii}$ | 60.0 |
| $S^{v}$—Mo—$Mo^{iii}$ | 90.0 | $S^{i}$—Mo—$Mo^{ix}$ | 90.0 |
| $S^{i}$—Mo—$Mo^{vi}$ | 131.89 (11) | $S^{ii}$—Mo—$Mo^{ix}$ | 48.11 (11) |
| $S^{ii}$—Mo—$Mo^{vi}$ | 48.11 (11) | $S^{iii}$—Mo—$Mo^{ix}$ | 131.89 (11) |
| $S^{iii}$—Mo—$Mo^{vi}$ | 131.89 (11) | S—Mo—$Mo^{ix}$ | 90.0 |
| S—Mo—$Mo^{vi}$ | 48.11 (11) | $S^{iv}$—Mo—$Mo^{ix}$ | 131.89 (11) |
| $S^{iv}$—Mo—$Mo^{vi}$ | 90.0 | $S^{v}$—Mo—$Mo^{ix}$ | 48.11 (11) |
| $S^{v}$—Mo—$Mo^{vi}$ | 90.0 | $Mo^{iii}$—Mo—$Mo^{ix}$ | 120.0 |
| $Mo^{iii}$—Mo—$Mo^{vi}$ | 180.0 | $Mo^{vi}$—Mo—$Mo^{ix}$ | 60.0 |
| S$^{i}$—Mo—Mo$^{v}$ | 48.11 (11) | Mo$^{v}$—Mo—Mo$^{ix}$ | 60.0 |
| S$^{ii}$—Mo—Mo$^{v}$ | 90.0 | Mo$^{vii}$—Mo—Mo$^{ix}$ | 120.0 |
| S$^{iii}$—Mo—Mo$^{v}$ | 90.0 | Mo$^{viii}$—Mo—Mo$^{ix}$ | 180.0 |
| S—Mo—Mo$^{v}$ | 131.89 (11) | Mo—S—Mo$^{vii}$ | 83.8 (2) |
| S$^{iv}$—Mo—Mo$^{v}$ | 131.89 (11) | Mo—S—Mo$^{vi}$ | 83.8 (2) |
| S$^{v}$—Mo—Mo$^{v}$ | 48.11 (11) | Mo$^{vii}$—S—Mo$^{vi}$ | 83.8 (2) |
| Mo$^{iii}$—Mo—Mo$^{v}$ | 60.0 | | |

Symmetry codes: (i) -x, -y, -z; (ii) -x+1, -y+1, -z; (iii) x-1, y-1, z; (iv) -x, -y+1, -z; (v) x, y-1, z; (vi) x+1, y+1, z; (vii) x, y+1, z; (viii) x-1, y, z; (ix) x+1, y, z.

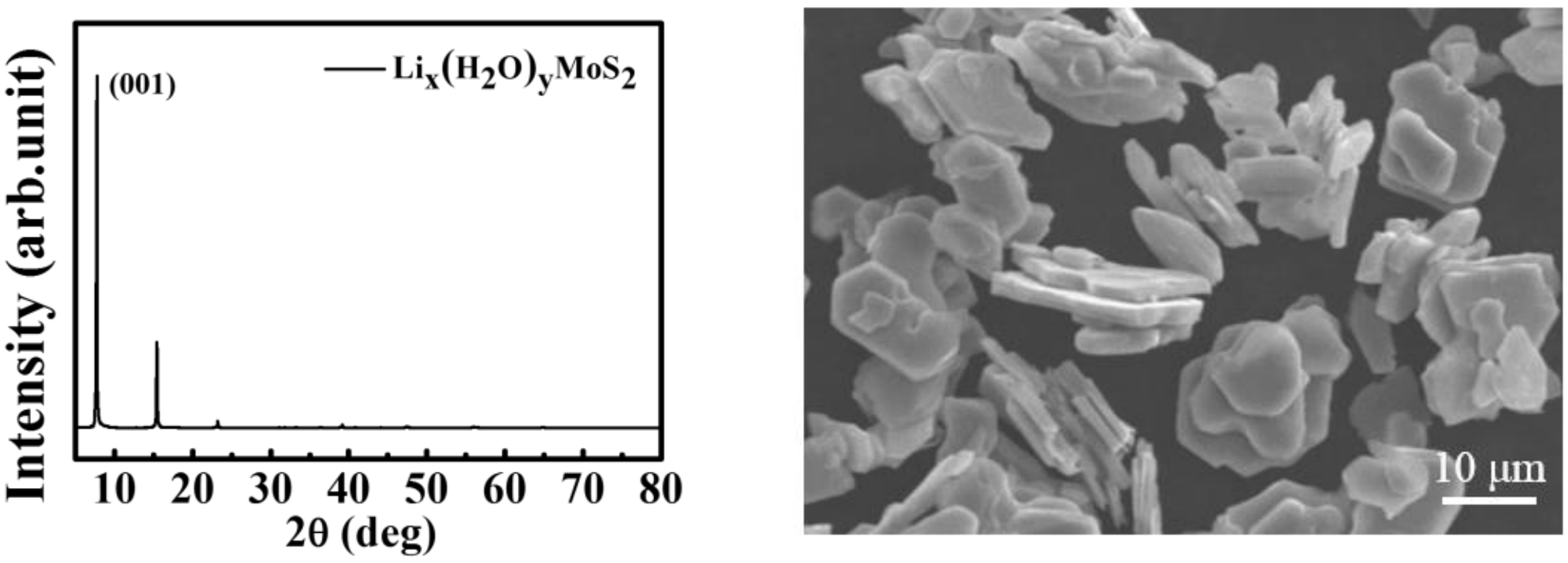

**Figure S1** (a) The X-ray diffraction of $Li_x(H_2O)_yMoS_2$. (b) The SEM image of the bulk 1T $MoS_2$

**Table S5.** The result of the EDX measurements.

| Spots | Atomic ratio (%) | |
|---|---|---|
| | Mo | S |
| 1 | 32.54 | 67.46 |
| 2 | 31.45 | 68.55 |
| 3 | 34.51 | 65.49 |
| 4 | 33.31 | 66.69 |
| 5 | 33.34 | 66.66 |
| Average | 33.03 | 66.97 |

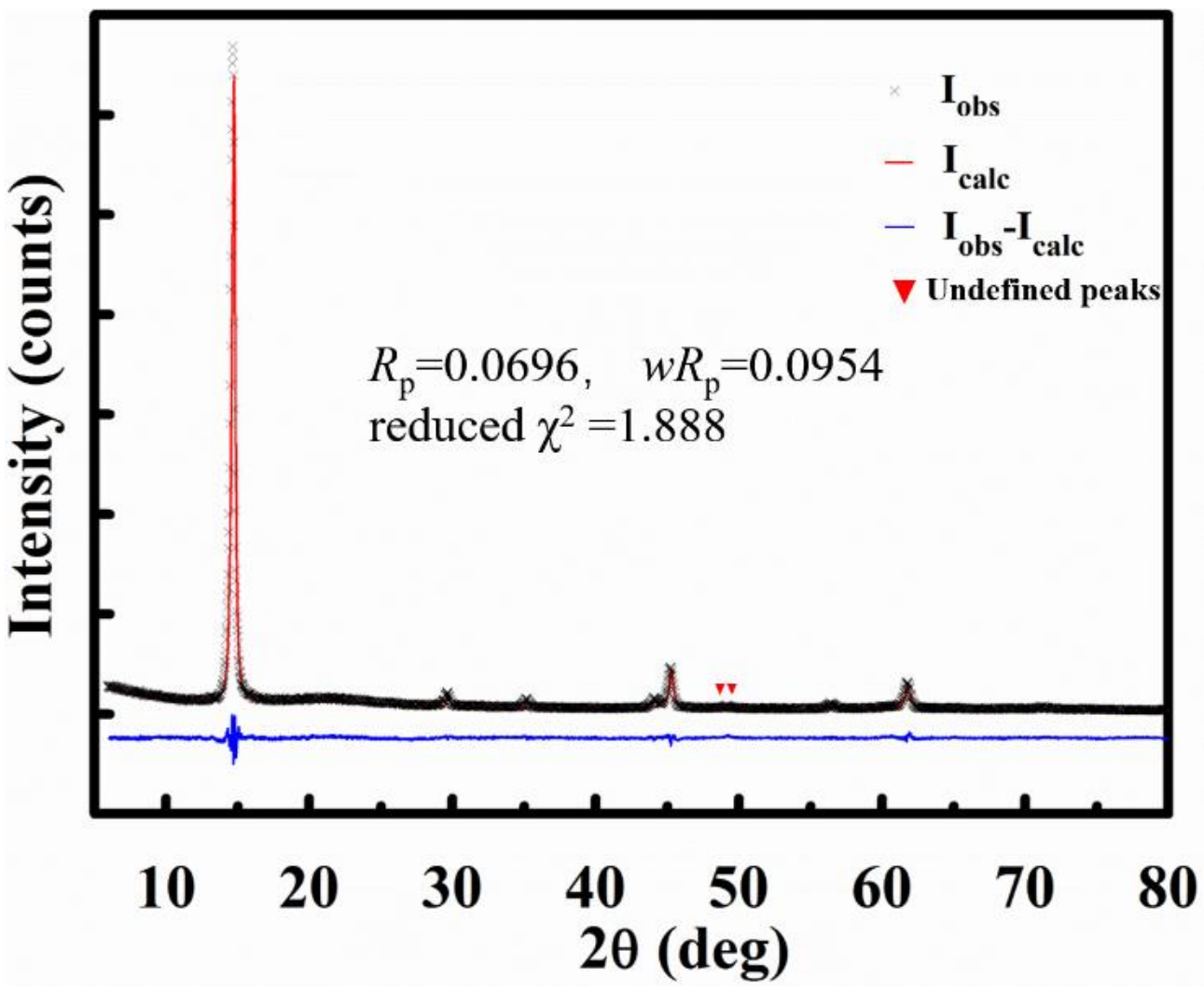

**Figure S2.** The Rietveld-Refinement of the PXRD data of bulk 1T $MoS_2$.

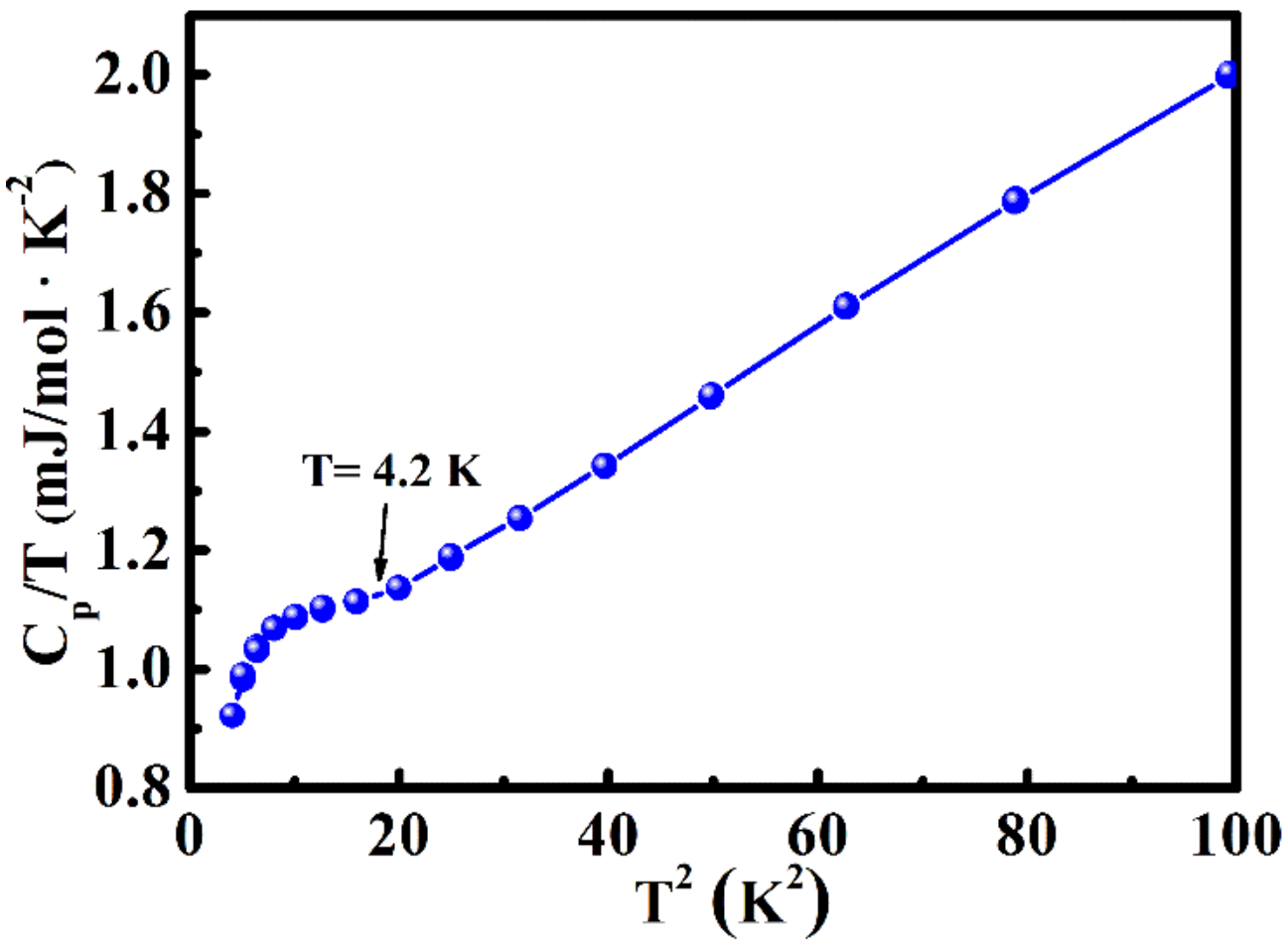

**Figure S3.** The specific heat capacity of bulk 1T $MoS_2$.

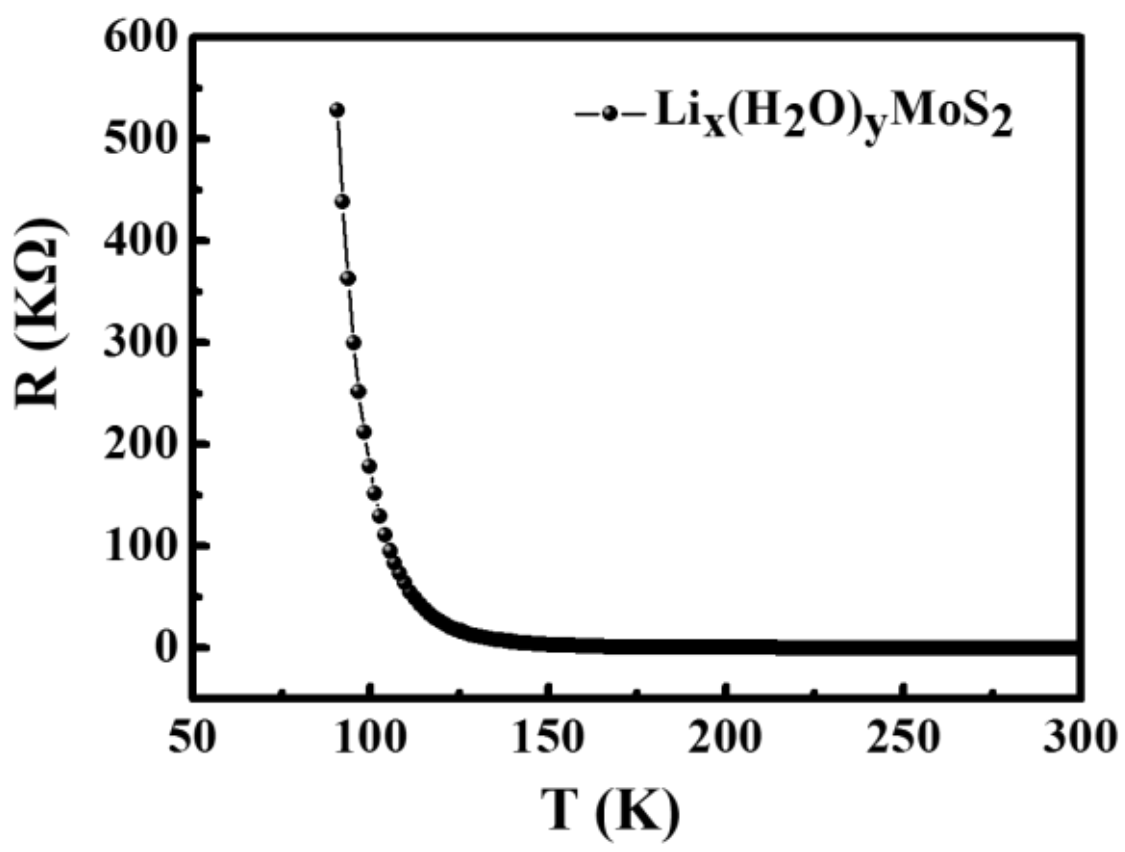

**Figure S4.** The temperature dependence of electrical resistivity of $Li_{1-x}(H_2O)_yMoS_2$.